\documentclass[twocolumn,showpacs,preprintnumbers,amsmath,amssymb,floatfix]{revtex4}
\usepackage{epsfig}
\usepackage{wasysym}
\usepackage{amssymb}
\usepackage{slashed}
\usepackage{physics}
\usepackage{color}

\begin{document}

\newcommand{\bvec}[1]{\mbox{\boldmath ${#1}$}}
\title{Pure spin-3/2 representation with consistent interactions}
\author{T. Mart}\email[]{terry.mart@sci.ui.ac.id}
\author{J. Kristiano}
\author{S. Clymton}
\affiliation{Departemen Fisika, FMIPA, Universitas Indonesia, Depok 16424, 
  Indonesia}
\date{\today}
\begin{abstract}
  We have investigated the use of pure spin-3/2 propagator with consistent interaction
  Lagrangians to describe the property of spin-3/2 resonance. 
  For this purpose we use the antisymmetric tensor spinor representation.
  By using the primary and secondary constraints we obtain the interaction 
  fields that have the correct degrees of freedom. To visualize the result 
  we calculate the contribution of spin-3/2 $\Delta$ resonance to the total 
  cross section of pion scattering and pion photoproduction off
  the nucleon. The result confirms that the scattering and photoproduction 
  amplitudes obtained from the pure spin-3/2 representation with
  consistent interaction Lagrangians exhibit the required property of a resonance.
  Therefore, the formalism can be used for phenomenological investigations in
  the realm of nuclear and particle physics.
\end{abstract}

\maketitle

\section{Introduction}
In the nuclear and particle physics the formulation of spin-3/2 particle 
constitutes an arduous and long-standing problem. So far, such a particle 
is commonly represented by the Rarita-Schwinger (RS) field \cite{rarita}, 
which is described by the tensor product of a vector   ${\textstyle 
\left(\frac{1}{2},\frac{1}{2} \right)}$ and a Dirac ${\textstyle \left(\frac{1}{2},0 \right) 
\oplus \left(0,\frac{1}{2} \right)}$  fields. Mathematically, the result of this product 
is well known, i.e., $\left(1,\frac{1}{2} \right) 
\oplus \left(\frac{1}{2},1 \right) 
\oplus  \left(\frac{1}{2},0 \right) 
\oplus \left(0,\frac{1}{2} \right)$
 \cite{Weinberg:1995mt}, 
which exhibits that the RS field contains two fields; 
the ${\textstyle \left(1,\frac{1}{2} \right) 
\oplus \left(\frac{1}{2},1 \right)}$ and the Dirac fields. 
The latter can be  eliminated by using an orthogonality relation
and as a result we obtain a spin 3/2 field that simultaneously contains
a spin 1/2 background; the  ${\textstyle \left(1,\frac{1}{2} \right) 
\oplus \left(\frac{1}{2},1 \right)}$ field.

The RS field has also another fundamental problem called 
the Velo-Zwanziger problem \cite{Velo-Zwanziger}. This problem originates
from the non-causal propagation of the wave front when the derivative terms 
of the RS field is gauged with the electromagnetic field. It was shown that 
the Velo-Zwanziger problem is related to the violation of constraints \cite{hasumi}. 
The interaction of spin-3/2 field with other fields should be constructed to 
have the same symmetry as the free field Lagrangian in order to preserve 
the correct degrees of freedom. For example, the earliest version of the 
$\pi N \Delta$ coupling, that has an off-shell parameter \cite{Benmerrouche:1989uc}, 
does not posses the local symmetry of RS field 
\cite{pascalutsa_1999}. Such a problem could be solved by introducing the 
gauge-invariant (GI) interaction to decouple the unphysical spin-1/2 background 
from the calculated transition amplitude \cite{pascalutsa_1999}.

Actually, the formalism of spin-3/2 particle can be presented by the 
pure spin-3/2 field ${\textstyle \left(\frac{3}{2},0 \right) 
\oplus \left(0,\frac{3}{2} \right)}$, which is clearly free from the 
spin-1/2 background. However, the problem with this field is that 
it uses 8-dimensional spinor since the spin-3/2 operator is represented 
by $4 \times 4$ matrices. Whereas the free 8-dimensional field has been
formulated by Weinberg \cite{weinberg}, it was still intricate to construct the 
corresponding interaction Lagrangian due to the non-covariant form of the 
8-dimensional field, until Acosta {\it et al.}  \cite{DelgadoAcosta:2015ypa} 
could embed the pure spin-3/2 field into a totally antisymmetric tensor of 
second rank. Since the components of the tensor are spinor, such representation 
is called the antisymmetric tensor spinor (ATS). The ATS representation is formed 
by a tensor product of antisymmetric field and Dirac spinor
\begin{eqnarray}
&& {\textstyle
\left[ \left(1,0 \right) \oplus \left(0,1 \right) \right] 
\otimes\left[ \left(\frac{1}{2},0 \right) \oplus \left(0,\frac{1}{2} \right) 
\right] ~=~}
\nonumber\\ 
& & {\textstyle
\left[ \left(\frac{3}{2},0 \right) \oplus \left(0,\frac{3}{2} 
\right) \right] \oplus \left[ \left(1,\frac{1}{2} \right) \oplus 
\left(\frac{1}{2},1 \right) \right]}
\nonumber\\ && 
{\textstyle
 \oplus \left[ \left(\frac{1}{2},0 \right) \oplus \left(0,\frac{1}{2} 
\right) \right] .}
\end{eqnarray}
In the ATS representation the pure spin-3/2 field is projected out by 
the Lorentz projection operator.

In the previous paper we have briefly reported the use of pure spin-3/2 
propagator to describe the $\Delta$ resonance in the $\pi N$ scattering 
\cite{kristiano2017}. It was shown that the conventional GI interaction 
Lagrangian cannot describe the resonance behavior of the spin-3/2 $\Delta$
baryon, unless the interaction was modified by adding an extra momentum 
dependence. Obviously, there was a lack of theoretical basis to support this
solution. Furthermore, the theoretical consistency of such 
an adhoc interaction Lagrangian could be questioned. 
In this paper we present a complete result of our investigation on 
the pure spin-3/2 formalism. We first discuss the ATS formalism and
its problem in describing the properties of a resonance. In 
Ref.~\cite{DelgadoAcosta:2015ypa} this problem was not observed
since the proposed phenomenological application is Compton scattering,
in which the spin-3/2 particle is on-shell and does not resonate. 
For the sake of simplicity, we choose the $\pi$-$N$ scattering to 
visualize the present problem. Then, we search for the 
consistency requirement in the interaction Lagrangians and present 
an example of consistent interaction Lagrangians for hadronic and 
electromagnetic interactions. By using these Lagrangians and the pure
spin-3/2 propagator we show that this formalism can be used for the purpose
of phenomenological applications.

We organize this paper as follows. In Sec.~\ref{sec:problem} we present the 
formalism of ATS and the corresponding problem to describe the resonance
properties. In Sec.~\ref{sec:consistent} we explain the construction of 
consistent interaction Lagrangians. Section \ref{sec:result} exhibits the 
numerical result and visualization of the resonance behavior of the pure 
spin-3/2 representation in the pion scattering and pion photoproduction 
processes. Finally, in Sec.~\ref{sec:conclusion} we summarize our 
investigation and conclude our findings.

\section{ATS and its problem}
\label{sec:problem}
In what follows we briefly summarize the ATS formalism and show that this
formalism has a problem to describe the properties of a resonance. We have
discussed this topic in our previous Rapid Communication \cite{kristiano2017}.
Let us start with the Casimir operator
$F = {\textstyle\frac{1}{4}} J_{\mu \nu} J^{\mu \nu}$ with $J^{\mu\nu}$ 
the angular momentum operator. For the field $\ket{(a,b)}$ this Casimir operator
has the eigenvalue equation
\begin{equation}
F \ket{(a,b)} = C(a,b) \ket{(a,b)},
\end{equation}
with the eigenvalue $C(a,b) = a(a+1) + b(b+1)$. By using this Casimir operator we can
construct the projection operator that can remove the $(1,{\textstyle\frac{1}{2}}) 
\oplus ({\textstyle\frac{1}{2}},1)$ and $({\textstyle\frac{1}{2}},0) \oplus 
(0,{\textstyle\frac{1}{2}})$ fields from the ATS. The projection operator reads
\begin{equation}
\mathcal{P} = \frac{[F - C(1,{\textstyle\frac{1}{2}})] [(F - C({\textstyle\frac{1}{2}},0)]}{[C({\textstyle\frac{3}{2}},0) - C(1,{\textstyle\frac{1}{2}})] [C({\textstyle\frac{3}{2}},0) - C({\textstyle\frac{1}{2}},0)]} ~.
\end{equation}
Acosta {\it et al.} \cite{DelgadoAcosta:2015ypa} have shown that this projection 
operator can be written as
\begin{equation}
\mathcal{P}_{\alpha \beta \gamma \delta} = {\textstyle\frac{1}{8}} \left( \sigma_{\alpha \beta} \sigma_{\gamma \delta} + \sigma_{\gamma \delta} \sigma_{\alpha \beta} \right) - {\textstyle\frac{1}{12}} \sigma_{\alpha \beta} \sigma_{\gamma \delta} \, ,
\end{equation}
with 
\begin{equation}
\sigma_{\alpha \beta} = {\textstyle\frac{i}{2}}\left[ \gamma_\alpha, \gamma_\beta \right]\, .
\end{equation}
This projection operator assures that the ATS formalism has  only the 
$({\textstyle\frac{3}{2}},0) \oplus (0,{\textstyle\frac{3}{2}})$ representation. 
In the ATS representation the pure spin-3/2 spinor is obtained by 
operating a pure spin-3/2 projection operator to the GI RS spinor, i.e.,
\cite{DelgadoAcosta:2015ypa}
\begin{equation}
w^{\mu \nu}({\bf p}, \lambda) = 2{\mathcal{P}^{\mu \nu}}_{\alpha \beta} U^{\alpha \beta}({\bf p}, \lambda) \,,
\end{equation}
where $\lambda = -{\textstyle\frac{3}{2}}, -{\textstyle\frac{1}{2}}, +{\textstyle\frac{1}{2}}, 
+{\textstyle\frac{3}{2}}$ are the $z$-components of the 
spin-3/2 operator eigenvalues and $U^{\alpha \beta}({\bf p}, \lambda)$ is 
the GI RS spinor, given by
\begin{equation}
U^{\alpha \beta}({\bf p}, \lambda) = \frac{1}{2m} \left[ p^\alpha \mathcal{U}^\beta({\bf p}, \lambda) - p^\beta \mathcal{U}^\alpha({\bf p}, \lambda)  \right],
\label{eq:U_ab_GI_spinor}
\end{equation}
with $\mathcal{U}^\alpha({\bf p}, \lambda)$ the RS vector-spinor. 
Clearly, except for the normalization constant $(2m)^{-1}$, the GI 
RS spinor $U^{\alpha \beta}({\bf p}, \lambda)$ given in
Eq.~(\ref{eq:U_ab_GI_spinor}) is identical to the GI RS field 
tensor $\Delta^{\mu \nu} = \partial^\mu 
\Delta^\nu - \partial^\nu \Delta^\mu$ given in 
Ref.~\cite{pascalutsa_1999}. 
Therefore, the difference between the ATS and the GI RS  representations 
is in their projection operators. The ATS projection operator is completely 
different from the common projection operator in RS field. The former
projects out the $({\textstyle\frac{3}{2}},0) \oplus (0,{\textstyle\frac{3}{2}})$ 
field, whereas the latter projects out the $(1,{\textstyle\frac{1}{2}}) \oplus 
({\textstyle\frac{1}{2}},1)$ field. 

In the pure spin-3/2 representation the corresponding propagator can be written as 
\cite{DelgadoAcosta:2015ypa}
\begin{equation}
S_{\alpha \beta \gamma \delta}(p) = \frac{\Delta_{\alpha \beta \gamma \delta}(p)}{p^2 - m^2 + i\epsilon} \, ,
\label{eq:propagator_pure}
\end{equation}
where 
\begin{equation}
\Delta_{\alpha \beta \gamma \delta}(p) = \left( \frac{p^2}{m^2} \right) \mathcal{P}_{\alpha \beta \gamma \delta} - \left( \frac{p^2-m^2}{m^2} \right) 1_{\alpha \beta \gamma \delta} \, ,
\label{eq:purepropagator}
\end{equation}
and $1_{\alpha \beta \gamma \delta}$ is the identity in the ATS space, i.e.,
\begin{equation}
1_{\alpha \beta \gamma \delta} = {\textstyle\frac{1}{2}} \left( g_{\alpha \gamma}g_{\beta \delta} - g_{\alpha \delta}g_{\beta \gamma} \right)1_{4 \times 4} \, .
\end{equation}
By using the orthogonality relation for the projection operator 
$\gamma^\mu \mathcal{P}_{\mu \nu \rho \sigma} = 0$, one may easily prove that the pure 
spin-3/2 spinor satisfies  $\gamma_\mu w^{\mu \nu}({\bf p}, \lambda) = 0$. 
This relation can be used to reduce the number of degrees of freedom 
(DOF) in the ATS representation, i.e.,
$6 \times 4 = 24$, by  $4 \times 4 = 16$. As expected, the pure spin-3/2 field 
in the ATS representation has $24-16=8$ DOF. 

Finally, for the purpose of the phenomenological application, such as meson-nucleon
scattering, it is important to note that 
the free Lagrangian for the pure spin-3/2 field in the ATS representation 
can be written as \cite{DelgadoAcosta:2015ypa}
\begin{equation}
\mathcal{L} = (\partial^\mu \Psi^{\alpha \beta}) \Gamma_{\mu \nu \alpha \beta \gamma \delta} (\partial^\nu \Psi^{\gamma \delta}) - m^2 \Psi^{\mu \nu} \Psi_{\mu \nu} \, ,
\label{eq:free_Lagrangian}
\end{equation}
where 
\begin{equation}
\Gamma_{\mu \nu \alpha \beta \gamma \delta} = 	4 g^{\sigma \rho} \mathcal{P}_{\alpha \beta \rho \mu} \mathcal{P}_{\sigma \nu \gamma \delta} \, ,
\end{equation}
and $\Psi^{\mu \nu}$
is the $({\textstyle\frac{3}{2}},0) \oplus (0,{\textstyle\frac{3}{2}})$ field. 
The kinetic term of the Lagrangian is invariant 
under the following gauge transformation
\begin{equation}
\Psi^{\mu \nu} \rightarrow \Psi^{\mu \nu} + \xi^{\mu \nu} \, ,
\label{eq:gaugetrans}
\end{equation}
where the antisymmetric tensor $\xi^{\mu \nu}$ is given by 
\begin{equation}
\xi^{\mu \nu} = \gamma^\mu \partial^\nu \xi - \gamma^\nu \partial^\mu \xi \, .
\end{equation}

\begin{figure}[t]
  \begin{center}
    \leavevmode
    \epsfig{figure=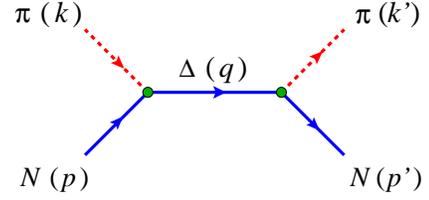,width=55mm}
    \caption{(Color online) Feynman diagram for the elastic $\pi N$ 
      scattering with a $\Delta$ resonance in the intermediate state.
      }
   \label{fig:feynman_piN} 
  \end{center}
\end{figure}

As stated in the Introduction we have found that the ATS formalism has a problem 
in describing the properties of a resonance. To explain this problem let us 
consider the elastic $\pi N$ scattering with a $\Delta$ resonance in the 
intermediate state. The corresponding Feynman diagram is displayed 
in Fig.~\ref{fig:feynman_piN}, in which the momenta of all involved particles are
shown for our convention. In the literature we note that the popular choice of 
Lagrangian for the $\pi N \Delta$ interaction reads \cite{pascalutsa_1999}
\begin{equation}
\mathcal{L}_{\pi N \Delta} = \left( \frac{g_{\pi N \Delta}}{m_\pi} \right) 
\bar{\Delta}^\mu \Theta_{\mu \nu}(z) N \partial^\nu \pi + {\rm H.c.} \,,
\end{equation}
where $\Delta^\mu$, $N$, and $\pi$  denote the 
$\Delta$-baryon vector-spinor, nucleon spinor, and pion field, respectively.
The tensor $\Theta_{\mu \nu}(z)$ is given by
\begin{equation}
\Theta_{\mu \nu}(z) = g_{\mu \nu} - \left(  z+{\textstyle\frac{1}{2}} \right) \gamma_\mu \gamma_\nu \, .
\label{eq:theta_mu_nu}
\end{equation}
Note that the constant $z$ in Eq.~(\ref{eq:theta_mu_nu}) is arbitrary and 
conventionally called the off-shell parameter. 
As stated before this Lagrangian does not posses any local symmetries of 
the RS field, and as a consequence it induces the unphysical lower-spin 
DOF, which is called spin 1/2 background \cite{pascalutsa_1999}. 
To decouple this unphysical background from the $\Delta$-exchange amplitude 
Pascalutsa and Timmermans introduce a GI interaction, which is given by \cite{pascalutsa_1999}
\begin{equation}
\mathcal{L}_{\pi N \Delta} = \left( \frac{g_{\pi N \Delta}} {m_\pi m_\Delta} \right) 
\bar{N} \gamma_5 \gamma_\mu \tilde{\Delta}^{\mu \nu}\partial_\nu \pi + {\rm H.c.} \,,
\end{equation}
where $\tilde{\Delta}^{\mu \nu}$ is the dual tensor of GI 
RS field tensor $\Delta^{\mu \nu}$. The latter is given by 
\begin{equation}
\Delta_{\mu \nu} = \partial_\mu \Delta_\nu - \partial_\nu \Delta_\mu \, .
\end{equation}
This GI interaction yields the $\Delta$-exchange amplitude
\begin{equation}
\Gamma^\mu (k') S_{\mu \nu}(q) \Gamma^\nu (k) = \frac{(g_{\pi N \Delta}/m_\pi)^2}{\slashed{q} 
- m_\Delta} \frac{q^2}{m_\Delta^2} P_{\mu \nu}^{(3/2)}(q) k'^\mu k^\nu \, ,
\label{eq:GI_interaction}
\end{equation}
with $P_{\mu \nu}^{(3/2)}$ the spin-3/2 projection operator in the RS field, i.e.,
\begin{equation}
P_{\mu \nu}^{(3/2)}(q) = g_{\mu\nu} - {\textstyle\frac{1}{3}} \gamma_\mu \gamma_\nu - \frac{1}{3q^2} (\slashed{q} \gamma_\mu q_\nu + q_\mu \gamma_\nu \slashed{q}) \, .
\end{equation}

In analogy to the GI interaction described above we can also construct the 
$\pi N \Delta$ interaction in the ATS formalism by changing the GI RS field 
tensor to the $({\textstyle\frac{3}{2}},0) \oplus (0,{\textstyle\frac{3}{2}})$ representation,
\begin{equation}
\mathcal{L}_{\pi N \Delta} = g_{\pi N \Delta}
\bar{N} \gamma_5 \gamma_\mu \tilde{\Psi}^{\mu \nu}\partial_\nu \pi + {\rm H.c.} \, ,
\label{eq:L_piND}
\end{equation}
where $\Psi^{\mu\nu}$ is the $({\textstyle\frac{3}{2}},0) \oplus (0,{\textstyle\frac{3}{2}})$ 
field tensor and $\tilde{\Psi}^{\mu \nu}$ is its dual tensor. In terms of vertex factor used
in many phenomenological applications \cite{Clymton:2017nvp,Mart:2015jof}, the interaction Lagrangian
given in Eq.~(\ref{eq:L_piND}) can be translated as
\begin{equation}
\Gamma_{\mu\nu}(k) = g_{\pi N \Delta} \gamma_5 \gamma_\mu k_\nu \, .
\label{eq:vertex_pure1}
\end{equation}
Thus, the $\Delta$-exchange amplitude in the ATS formalism can be written as 
$\Gamma_{\mu\nu}(k') \tilde{S}^{\mu \nu \rho \sigma}(q) \Gamma_{\rho \sigma}(k)$, 
where $\tilde{S}^{\mu \nu \rho \sigma}$ is defined by 
\begin{eqnarray}
\Gamma_{\mu\nu}(k') \tilde{S}^{\mu \nu \rho \sigma}(q) \Gamma_{\rho \sigma}(k) &=& 
{\textstyle\frac{1}{4}} g_{\pi N \Delta}^2 \epsilon^{\mu \nu \alpha \beta} 
\epsilon^{\rho \sigma \kappa \lambda} \gamma_5 \gamma_\mu \nonumber\\
&\times& S_{\alpha \beta \kappa \lambda}(q) \gamma_\rho \gamma_5 k'_\nu k_\sigma \, .
\label{eq:delta_exchange}
\end{eqnarray} 
By using Eqs.~(\ref{eq:propagator_pure}) and (\ref{eq:vertex_pure1}) we 
can directly calculate Eq.~(\ref{eq:delta_exchange}) and it is easy
to show that the nonvanishing amplitude is only the term obtained from 
the contraction with the identity $1_{\alpha \beta \kappa \lambda}$, 
since on the right hand side of Eq.~(\ref{eq:delta_exchange}) 
the contraction with $\mathcal{P}_{\alpha \beta \kappa \lambda}$ 
vanishes due to the orthogonality relation 
$\gamma^\alpha \mathcal{P}_{\alpha \beta \kappa\lambda} = 0$ 
and the fact that $\tilde{\sigma}^{\mu \nu} = - \gamma_5 \sigma^{\mu \nu}$. 
By calculating this nonvanishing $\Delta$-exchange amplitude 
we obtain that 
\begin{eqnarray}
\Gamma_{\mu\nu}(k') \tilde{S}^{\mu \nu \rho \sigma}(q) \Gamma_{\rho \sigma} (k)
&=& \frac{g_{\pi N \Delta}^2 \left( q^2-m_\Delta^2 \right)}{m_\Delta^2 
(q^2 - m_\Delta^2 + i\epsilon)}\nonumber\\
&\times& \left(  g^{\nu \sigma} + 
{\textstyle\frac{1}{2}} \gamma^\nu \gamma^\sigma \right) k'_\nu k_\sigma \, . ~~
\label{eq:gammap_mu_nu}
\end{eqnarray}
Obviously, Eq.~(\ref{eq:gammap_mu_nu}) does not show the behavior of a resonance, 
since at the resonance pole ($q^2 = m_\Delta^2$) the amplitude is equal 
to zero, instead of being maximum. Thus, we may conclude that the 
interaction Lagrangian given by Eq.~(\ref{eq:L_piND}) cannot be used for calculating
the resonance contribution.

The source of problem is coming from the GI interaction Lagrangian given by 
Eq.~(\ref{eq:L_piND}), i.e., the contraction between gamma matrix and 
the projection operator of pure spin-3/2 field $\mathcal{P}_{\alpha \beta \kappa \lambda}$ 
vanishes. It is also obvious that this problem can be easily solved by 
modifying the interaction Lagrangian, e.g., by replacing the gamma matrix 
with a partial derivative, 
\begin{equation}
\mathcal{L}_{\pi N \Delta} = \left( \frac{g_{\pi N \Delta}}{m_\Delta} \right) 
\bar{N} \gamma_5 \partial^\mu \Psi_{\mu \nu}\partial^\nu \pi + {\rm H.c.} \, ,
\label{eq:LPiNDpartial}
\end{equation}
with the corresponding vertex factor
\begin{equation}
\Gamma^{\mu \nu}(k) = \left( \frac{g_{\pi N \Delta}}{m_\Delta} \right) \gamma_5 q^\mu k^\nu \, .
\label{eq:vertex_derivative}
\end{equation}
By using this vertex factor we can calculate 
the $\Delta$-exchange amplitude to obtain 
\begin{eqnarray}
&&\Gamma^{\mu \nu}(k') S_{\mu \nu \rho \sigma}(q) \Gamma^{\rho \sigma}(k) \nonumber\\
&=& \left( \frac{g_{\pi N \Delta}}{m_\Delta} \right)^2 \gamma_5 q^\mu 
 S_{\mu \nu \rho \sigma}(q) \gamma_5 q^\rho k'^\nu k^\sigma \nonumber\\
&=& \frac{g_{\pi N \Delta}^2 k'^\nu k^\sigma}{q^2 - m_\Delta^2 + i\epsilon} 
\left[ \frac{q^4}{4m_\Delta^4} P^{(3/2)}_{\nu \sigma}(q)  \right.
\nonumber\\
&-& \left. 
\left( \frac{q^2 - m_\Delta^2}{2m_\Delta^4} \right) \left(  q^2 g_{\nu \sigma} - q_\nu q_\sigma \right) \right],~~~~
\label{eq:amplitude_pure}
\end{eqnarray}
which is different from the result of GI interaction given by 
Eq.~(\ref{eq:GI_interaction}) by the second term.  This term 
is significant only at energies far beyond the resonance pole
and, as in Eq.~(\ref{eq:gammap_mu_nu}), does not show the 
property of a resonance. Nevertheless, the result given by 
Eq.~(\ref{eq:amplitude_pure}) is very interesting, because at 
the resonance pole, i.e., $q^2 = m_\Delta^2$, the second term vanishes and 
the $\Delta$-exchange amplitude is proportional to that obtained from
the GI RS interaction, i.e., Eq.~(\ref{eq:GI_interaction}).

To conclude this Section we may safely say that although for certain
types of interaction Lagrangians the ATS representation cannot exhibit
the property of a resonance required for use in phenomenological studies
of hadronic physics, we are still able to choose different interaction
to overcome this issue. However, it is obvious that such a solution does not
have a strong theoretical basis and it is also possible that the suitable interaction 
found in this way is not unique. Therefore, we need a systematic mechanism to determine
the  genuine interaction through a number of relevant constraints. This is
the topic of our discussion in the following Section.

\section{Construction of the consistent interaction Lagrangians}
\label{sec:consistent}

The problem of constraint in the interacting RS field can be solved by 
constructing the interaction that has the same symmetry as the free field 
one \cite{pascalutsa_2001}. To this end, it is essential to check the impact 
of the interaction Lagrangian on the constraint and to carry out the 
Dirac-Faddeev quantization \cite{pascalutsa_1998}. This procedure can be 
generally applied to the higher spin baryons such as nucleon resonances. 
In the case of the interaction terms that are separable from the free field, 
i.e., $\mathcal{L} = \mathcal{L}_{\mathrm{free}} + \mathcal{L}_{\mathrm{int}}$, 
such procedure can be easily carried out. However, in the case of the 
interaction terms that originate from the gauged free field, we again face the 
Velo-Zwanziger problem. Such a gauged spin-3/2 field represents the spin-3/2 
lepton undergoing electroweak interaction with photon, $W^\pm$ and 
$Z^0$ bosons. We note that the search for excited state of leptons was 
performed by the ATLAS collaboration \cite{atlas}. It has been proposed that 
the excited state of leptons have a spin-3/2 state \cite{abdullah}. To this end, 
we can use the pure spin-3/2 field to find the gauged electroweak Lagrangian 
of spin-3/2 lepton, as pure spin-3/2 field is free from the Velo-Zwanziger 
problem \cite{Velo-Zwanziger}.

We start again with the free Lagrangian for pure spin-3/2 in the ATS representation 
given by Eq.~(\ref{eq:free_Lagrangian}).
The conjugate momenta of the fields are given by 
\begin{eqnarray}
&& \bar{\pi}_{\gamma \delta} = \frac{\partial \mathcal{L}}{\partial(\partial^0 \psi^{\gamma \delta})} = (\partial^\mu \bar{\psi}^{\alpha \beta}) \Gamma_{\mu 0 \alpha \beta \gamma \delta} , \\
&& \pi_{\gamma \delta} = \frac{\partial \mathcal{L}}{\partial(\partial^0 \bar{\psi}^{\gamma \delta})} = \Gamma_{0 \nu \gamma \delta \alpha \beta} (\partial^\nu \psi^{\alpha \beta}) \, ,
\end{eqnarray}
and they can be expressed as functions of the field "velocity" $\bar{v}^{\alpha \beta} 
= \partial^0 \bar{\psi}^{\alpha \beta}$ and $v^{\alpha \beta} = \partial^0 \psi^{\alpha \beta}$,
i.e.,
\begin{eqnarray}
\bar{\pi}_{\mu \nu} &=& \bar{v}^{\alpha \beta} {{\mathcal{P}_{\alpha \beta}}^\rho}_0 (g_{\mu \rho} \gamma_\nu - g_{\nu \rho} \gamma_\mu) \gamma_0 + \nonumber\\ 
&& (\partial^i \bar{\psi}^{\alpha \beta}) \Gamma_{i 0 \alpha \beta \mu \nu} \, ,
\end{eqnarray}
\begin{eqnarray}
\pi_{\mu \nu} &=& \gamma_0 (\gamma_\mu g_{\rho \nu} - \gamma_\nu g_{\rho \mu}) {\mathcal{P}^\rho}_{0 \alpha \beta} v^{\alpha \beta} + \nonumber\\
&& \Gamma_{0j \mu \nu \alpha \beta} (\partial^j \psi^{\alpha \beta}) \, .
\end{eqnarray}
Due to the non-invertible property of idempotent operator $P_{\alpha \beta \gamma \delta}$, 
the "velocity" cannot be expressed as a linear combination of conjugate momenta. 
The primary constraints arise from the condition that not all momenta are linearly independent, 
with the relations
\begin{equation}
\bar{\theta}_\rho^{(1)} = \bar{\pi}_{\rho \sigma} \gamma^\sigma, ~\theta_\rho^{(1)} = \gamma^\sigma \pi_{\rho \sigma} \, .
\end{equation}
Next, the Hamiltonian density of pure spin-3/2 field is given by 
\begin{equation}
\mathcal{H}_{3/2} = \bar{\pi}_{\gamma \delta} v^{\gamma \delta} + \bar{v}^{\gamma \delta} \pi_{\gamma \delta} - \mathcal{L} \, ,
\end{equation}
and the total Hamiltonian reads
\begin{eqnarray}
\mathcal{H}_T &=& \bar{\lambda}^\rho \theta_\rho^{(1)} + \bar{\theta}_\rho^{(1)} \lambda^\rho + \mathcal{H}_{3/2} \nonumber\\
&=& \bar{\lambda}^\rho \theta_\rho^{(1)} + \bar{\theta}_\rho^{(1)} \lambda^\rho - (\partial^i \bar{\psi}^{\alpha \beta}) \Gamma_{ij \alpha \beta \gamma \delta} (\partial^j \psi^{\gamma \delta}) \nonumber\\
&& - [\bar{\pi}_{\gamma \delta} - (\partial^i \bar{\psi}^{\alpha \beta}) \Gamma_{i0 \alpha \beta \gamma \delta}] \times \nonumber\\
&& [\pi^{\gamma \delta} - {{\Gamma_{0j}}^{\gamma \delta}}_{\alpha \beta} (\partial^j \psi^{\alpha \beta}) ] + m^2 \psi^{\mu \nu} \psi_{\mu \nu}\, .
\end{eqnarray}
The conditions of $\bigl\lbrace \bar{\theta}_\rho^{(1)}, \mathcal{H}_T \bigr\rbrace = 0$ 
and $\bigl\lbrace \theta_\rho^{(1)}, \mathcal{H}_T \bigr\rbrace = 0$ create the secondary constraints
\begin{equation}
\bar{\theta}_\rho^{(2)} = \bar{\psi}_{\rho \sigma} \gamma^\sigma, ~\theta_\rho^{(2)} = \gamma^\sigma \psi_{\rho \sigma} \, .
\end{equation}
Thus, the pure spin-3/2 fields end up with the secondary constraints, as the 
conditions of $\bigl\lbrace \bar{\theta}_\rho^{(2)}, \mathcal{H}_T \bigr\rbrace = 0$ 
and $\bigl\lbrace \theta_\rho^{(2)}, \mathcal{H}_T \bigr\rbrace = 0$ will be 
satisfied if the coefficients $\lambda^\rho$ and $\bar{\lambda}^\rho$ fulfill the relation
$4\lambda^\rho = \gamma^\rho \gamma_\mu \lambda^\mu$ and $4 \bar{\lambda}^\rho = 
\bar{\lambda}^\mu \gamma^\rho \gamma_\mu$.

In the case of massive pure spin-3/2 Lagrangian the number of DOF can be counted
as follows. Each of the fields $\psi^{\mu \nu}$ and its conjugate momentum $\pi^{\mu \nu}$ 
have $6 \times 4 = 24$ components, so in total they have 48 components. 
Each of the primary and secondary constraints reduces the DOF by $4 \times 4 = 16$ 
components. Hence, the number of independent components is $48-2\times 16=16$. 
This number describes the number of independent components 
of ${\textstyle \left(\frac{3}{2},0 \right) \oplus \left(0,\frac{3}{2} \right)}$ 
representation in phase space. Thus, each of the fields and its conjugate momentum 
have 8 independent components. For the massless case, the primary constraints become 
the first class constraints as the conditions of $\bigl\lbrace \bar{\theta}_\rho^{(1)}, 
\mathcal{H}_T \bigr\rbrace = 0$ and $\bigl\lbrace \theta_\rho^{(1)}, \mathcal{H}_T \bigr\rbrace = 0$ 
are identities. Therefore, the coefficients $\lambda^\rho$ and $\bar{\lambda}^\rho$ 
cannot be determined.

The general form of phase-space integral is
\begin{eqnarray}
Z &=& \int \mathcal{D}\psi^{\mu \nu} \mathcal{D}\bar{\psi}^{\mu \nu} \sqrt{\mathrm{det} \left\lbrace \theta_\alpha, \theta_\beta \right\rbrace} \prod_{n=1}^2 \delta(\theta_n) \delta(\bar{\theta}_n) \times \nonumber\\
&& \exp{i \int d^4x [\bar{\pi}_{\mu \nu} v^{\mu \nu} + \bar{v}^{\mu \nu} \pi_{\mu \nu} - \mathcal{H}_{3/2}]}\, .
\end{eqnarray}
The Poisson bracket of all constraint combinations are independent of the field. 
All constraints are proportional to the field or conjugate momentum. There are only two 
possibilities of the Poisson bracket of constraints, either zero or proportional to 
the gamma matrices. Therefore, the determinant factor is just a normalizing constant 
that can be ruled out from integral.

Generally, the interaction Lagrangian with pure spin-3/2 field can be written as
\begin{equation}
\mathcal{L} = \mathcal{L}_{3/2} + \bar{J}_{\mu \nu} \Psi^{\mu \nu} + \bar{\Psi}^{\mu \nu} J_{\mu \nu} \, .
\end{equation}
The interaction terms affect the secondary constraints, so the latter becomes
\begin{eqnarray}
&& \bar{\theta}_\rho^{(2)} = (m^2 \bar{\psi}_{\rho \sigma} - \bar{J}_{\rho \sigma}) \gamma^\sigma, \nonumber\\ 
&& \theta_\rho^{(2)} = \gamma^\sigma (m^2 \psi_{\rho \sigma} - J_{\rho \sigma})\, .
\end{eqnarray}
Because the interaction terms consist of other fields, we should constrain $J_{\mu \nu}$ 
in such a way that it will not affect the functional determinant of the constraints. 
To this end we pick the constraint as $ \gamma^\mu\, {J}_{\mu \nu} = 0$ and 
$\bar{J}_{\mu \nu} \gamma^\nu = 0$. 
In the pure spin-3/2 field formalism, incidentally, 
the projection operator $\mathcal{P}_{\mu \nu \rho \sigma}$ has the orthogonality 
relation $\gamma^\mu \mathcal{P}_{\mu \nu \rho \sigma} = 0$ and  
$\mathcal{P}_{\mu \nu \rho \sigma} \gamma^\rho = 0$.
As a consequence, the interaction Lagrangian could contain such projection operator 
and for the simplest consistent $\pi N\Delta$ interaction Lagrangian we have
\begin{equation}
\mathcal{L}_{\pi N \Delta} = \left( \frac{g_{\pi N \Delta}}{m_\Delta} \right) 
\bar{N} \gamma_5 \mathcal{P}_{\mu \nu \rho \sigma} \partial^\rho \psi^{\mu \nu} \partial^\sigma \pi + {\rm H.c.} \, ,
\label{eq:consistent_piN}
\end{equation}
whereas for the electromagnetic transition the corresponding 
$\gamma N \Delta$ Lagrangian  reads 
\begin{equation}
\mathcal{L}_{\gamma N \Delta} = \bar{N} (g_1 \mathcal{P}_{\rho \sigma \mu \nu}  + g_2 \gamma_\rho \mathcal{P}_{\sigma \alpha \mu \nu} \partial^\alpha) \psi^{\mu \nu} F^{\rho \sigma} + \mathrm{H. c.}
\label{eq:Lagrangian_em}
\end{equation}

\section{Numerical result and visualization}
\label{sec:result}

As in the previous report \cite{kristiano2017} we can explore
the behavior of the pure spin-3/2 propagator by 
comparing the $\Delta(1232)$ resonance contribution 
to the total cross section of elastic $\pi N$ scattering, 
obtained from the pure spin-3/2 propagator and the conventional ones. 
For this purpose it is important to 
include the resonance width $\Gamma$ in the resonance 
propagator, i.e., by replacing $i\epsilon\to i\Gamma m_\Delta$.

The elastic $\pi N$ scattering amplitude is traditionally written as
\begin{eqnarray}
\mathcal{M}&=&\bar{u}(p',s')(A+B\slashed{Q})u(p,s)\, ,
\label{eq:amplitude_M}
\end{eqnarray}
with $Q=(k+k')/2$. 
By using the RS propagator with GI interaction the amplitude obtained
from the Feynman diagram depicted in Fig.~\ref{fig:feynman_piN}, i.e.,
Eq.~(\ref{eq:GI_interaction}), can be decomposed into
\begin{eqnarray}
A &=& G_1 \left\lbrace m_N\left(3k' \cdot k - 2 p \cdot k - m_\pi^2 - {2}q \cdot k' q \cdot k/{q^2} \right)\right.\nonumber\\
& & \left. + m_\Delta\left(3k'\cdot k-2p\cdot k-m_\pi^2 -{2}m_\pi^2 q\cdot Q/{q^2}\right)\right\rbrace, ~~~~~ \\
B &=& G_1 \left\lbrace 3k' \cdot k -m_\pi^2 +2m_N^2 -{2} q \cdot k' q \cdot k/{q^2} \right.\nonumber\\
& & \left. +q\cdot (k'-k) + 2m_\Delta m_N \left( 1 - {q \cdot Q}/{q^2}\right)\right\rbrace \, ,
\end{eqnarray}
with 
\begin{eqnarray}
G_1 &=& {q^2\,g_{\pi N\Delta}^2}/[{3m_\pi^2 m_\Delta^2(q^2-m_\Delta^2 +i\Gamma m_\Delta)}] \, .
\end{eqnarray}

In the case of pure spin-3/2 propagator with the hadronic vertex given by
Eq.~(\ref{eq:vertex_derivative}), i.e., Eq.~(\ref{eq:amplitude_pure}), we obtain
\begin{eqnarray}
A &=& G_2\Bigl[ ({q^4}/{12m_\Delta^4}) \Bigl(3k' \cdot k - m_\pi^2 - 2 p \cdot k \nonumber\\
&& - 2 m_\pi^2\, q \cdot Q/{q^2} \Bigr)  - \left\{ (q^2 - m_\Delta^2)/2 m_\Delta^4\right\}\nonumber\\
& & \times (q^2 k' \cdot k - q \cdot k q \cdot k')\Bigr]\, ,~~~\\
B &=& \left({q^4 m_N G_2}/{6m_\Delta^4}\right) \left( 1 - {q \cdot Q}/{q^2}  \right) \, ,
\end{eqnarray}
with 
\begin{eqnarray}
G_2 &=& {g_{\pi N\Delta}^2}/[q^2 - m_\Delta^2 + i\Gamma m_\Delta] \, .
\end{eqnarray}

Finally, if we used the consistent interaction Lagrangian given by 
Eq.~(\ref{eq:consistent_piN}) we noted that the scattering amplitude
becomes more simple, i.e.,
\begin{eqnarray}
\mathcal{M}&=& G_3\,\bar{u}(p',s') \,\gamma_5 \,q_{\mu} k'_\nu\, \mathcal{P}^{\mu\nu\alpha\beta}\,
\gamma_5 \,q_{\alpha} k'_\beta\,u(p,s)\, ,~~ 
\label{eq:consistent_gN}
\end{eqnarray}
where
\begin{eqnarray}
G_3 &=& \frac{g_{\pi N\Delta}^2}{m_\Delta^2(s-m_\Delta^2 +i m_\Delta \Gamma_\Delta)} \, .
\end{eqnarray}
By decomposing Eq.~(\ref{eq:consistent_gN}) into Eq.~(\ref{eq:amplitude_M}) we obtain
\begin{eqnarray}
A &=& G_3 (q^2 m_\pi^2 - q\cdot k'\,^2)/6 \, ,\\
B &=& 0 \, .
\end{eqnarray}

By using the standard method \cite{bjorken} we can calculate 
the cross section from the scattering amplitude $\mathcal{M}$
given by Eq.~(\ref{eq:amplitude_M}). Note that the amplitudes 
obtained from the three different models are completely different.
For the sake of comparison we use different coupling constants
in order  to produce comparable total cross sections. Obviously, 
this would not raise a problem since almost all of the coupling 
constants in the phenomenological applications are fitted to 
reproduce the experimental data.

\begin{figure*}[t]
  \begin{center}
    \leavevmode
    \epsfig{figure=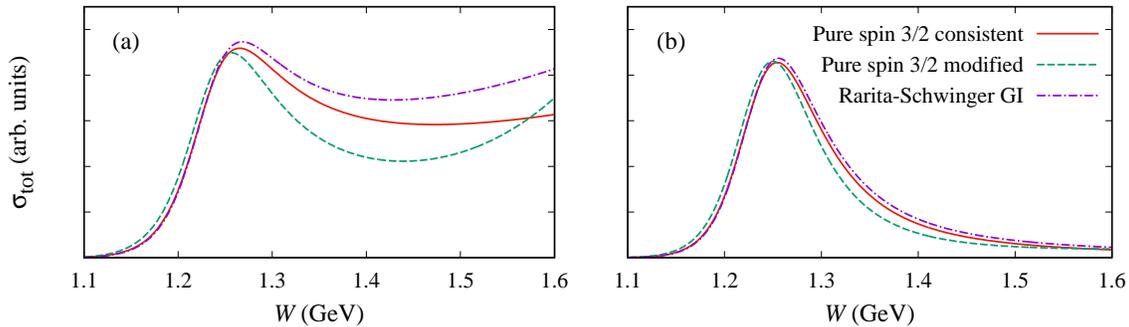,width=150mm}
    \caption{(Color online) Contribution of the $\Delta(1232)$ resonance 
      to the $\pi N\to \pi N$ scattering total cross section in arbitrary
      units (arb. units) calculated by using the pure spin-3/2 models with 
      consistent [Eq.~(\ref{eq:consistent_piN})] and  modified  
      [Eq.~(\ref{eq:LPiNDpartial})] interaction Lagrangians as well as the 
      Rarita-Schwinger model with GI interaction as 
      a function of total c.m. energy $W$. Panels (a) and (b) are obtained from
      the calculations without and with hadronic form factors in the 
      hadronic vertices, respectively. Note that to simplify the comparison we 
      do not use the same value of coupling constant in all calculations.
      }
   \label{fig:contribution_piN} 
  \end{center}
\end{figure*}

By taking point-particle approximation the total cross sections obtained
from the three models are depicted in panel (a) of Fig.~\ref{fig:contribution_piN}.
The resonance behavior centered around $W\approx 1.25$ GeV is produced by
all models, including the background phenomenon shown by the increase of
total cross section as the energy increases for $W\gtrsim 1.40$ GeV. 
The phenomenon originates from the momentum dependence in the numerator of 
Eq.~(\ref{eq:amplitude_pure}). The resonance background has another effect, i.e.,
shifting the resonance peak slightly from its original position at 1.232 GeV to higher 
energy. We also observe from panel (a) of Fig.~\ref{fig:contribution_piN} that
the backgrounds obtained from the pure spin-3/2 models are significantly smaller than
that of the RS model at $W\approx 1.40$ GeV. The reason can be traced back to the
second term in the square bracket of Eq.~(\ref{eq:amplitude_pure}). For 
 $W\gg 1.40$ GeV the first term of Eq.~(\ref{eq:amplitude_pure}) becomes dominant, 
since $q^4=W^4$, and the total contribution starts to diverge. From panel (a) of 
Fig.~\ref{fig:contribution_piN} it is also 
interesting to note that the pure spin-3/2 model with consistent interaction yields
moderate background compared to the other two models and does not show a divergence
behavior at high energies.

In the covariant Feynman diagrammatic approach the phenomenon of large background 
contribution is found to be natural. Alternatively, one can also interpret this
background as the contribution of a $Z$-diagram \cite{Jaroszewicz:1990mx}, i.e., 
a production of a particle and an antiparticle in the intermediate state not considered 
in Fig.~\ref{fig:feynman_piN}. Note that the phenomenon does not exist in the multipoles approach, 
where a perfect resonance structure can be produced by using the Breit-Wigner 
parameterization \cite{Mart:2017mwj}. 

The large background contribution can disturb the nature of the resonance 
itself and might induce other difficulties such as the problem to fit experimental data, 
especially in a covariant isobar model \cite{Clymton:2017nvp}, in which a large 
number of resonances are included while the individual resonance peaks are no 
longer distinguishable due to the proximity of their masses. To alleviate this 
problem it is customarily to use hadronic form factor (HFF) in each hadronic 
vertex shown in Fig.~\ref{fig:feynman_piN}. For a brief discussion of the 
HFF along with its problem with the gauge invariance we refer the reader to 
Refs.~\cite{Vrancx:2011qv,hbmf}. 

In spite of the objection that the HFF introduces
new free parameters, it should be noted that the existence of HFF is inevitable 
due to the fact that the baryon is not a point particle. Furthermore, the use of
HFF is also important to eliminate the divergence of the scattering amplitude.
Thus, in the present work we include the HFF and adopt the dipole HFF as in 
the previous work in the 
form of \cite{hbmf}
\begin{equation}
F = {\Lambda^4}/[\Lambda^4+\left( q^2-m_\Delta^2 \right)^2] \, ,
\label{eq:hff}
\end{equation}
where the hadronic cutoff is chosen to be $\Lambda=0.5$ GeV in order to produce
a reasonable resonance structure in the total cross section.
By including this HFF we obtain the result shown in panel (b) of 
Fig.~\ref{fig:contribution_piN}, in which a perfect resonance structure 
for all models is displayed. Compared to the pure spin-3/2 models the RS structure 
is slightly shifted to the right. This is understandable if we compare the
original contributions of all models (without HFF) as shown in panel (a) 
of Fig.~\ref{fig:contribution_piN}. Therefore, apart of its different formulation 
the pure spin-3/2 propagator still shows the usual resonance structure as in the
conventional RS propagator. Furthermore,  Fig.~\ref{fig:contribution_piN} clearly
indicates that to obtain the natural property of a resonance the use of HFF is 
mandatory in the covariant Feynman diagrammatic approach.

The next application of our present work is the contribution of
spin-3/2 $\Delta$ resonance to the pion photoproduction off a nucleon. As shown in the
previous work the choice of inappropriate electromagnetic interaction could fail to generate 
the correct property of a resonance in the cross section \cite{kristiano2017}. Thus,
in what follows we will calculate the $\Delta$ resonance contribution to the total 
cross section with a consistent interaction and compare the result with those of
previous works.

\begin{figure}[b]
  \begin{center}
    \leavevmode
    \epsfig{figure=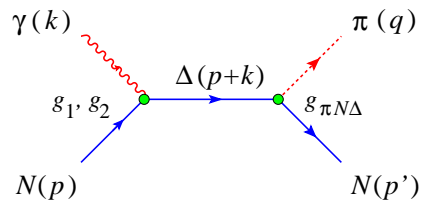,width=55mm}
    \caption{(Color online) Feynman diagram for the $\pi N$ photoproduction
      with a $\Delta$ resonance in the intermediate state. The electromagnetic
      and hadronic couplings are indicated in the diagram.
      }
   \label{fig:feynman_gN} 
  \end{center}
\end{figure}

The corresponding Feynman diagram is depicted in Fig.~\ref{fig:feynman_gN}. The hadronic 
vertex factor can be obtained from Eq.~(\ref{eq:consistent_piN}), i.e., 
\begin{eqnarray}
\Gamma_{\pi N \Delta}^{\mu\nu} &=& \frac{g_{\pi N\Delta}}{m_\Delta^2} \gamma_5 (p+k)_{\rho} q_\sigma \mathcal{P}^{\mu\nu\rho\sigma} \, ,
\label{eq:piNDelta}
\end{eqnarray}
whereas the electromagnetic one obtained from Eq.~(\ref{eq:Lagrangian_em}) can be written as
\begin{eqnarray}
\Gamma_{\gamma N \Delta}^{\alpha\beta} &=& \mathcal{P}^{\rho\sigma\alpha\beta} \Bigl\{g_1(k_\rho\epsilon_\sigma-\epsilon_\rho k_\sigma) \nonumber\\
&&+\frac{g_2}{m_\Delta} (p+k)_\sigma\,(\slashed{k}\epsilon_\rho- k_\rho\slashed{\epsilon})\Bigr\} ~.
\label{eq:gammaNDelta}
\end{eqnarray}
Note that in Eqs.~(\ref{eq:piNDelta}) and (\ref{eq:gammaNDelta}) we have inserted additional 
$m_\Delta$ in the denominator to make the coupling constants dimensionless. 
Furthermore, only two couplings are independent in this case, since the dual tensor
of ATS is proportional to the tensor itself, i.e.,
$\tilde{\Psi}_{\mu \nu} = -\gamma_5 \Psi_{\mu \nu} $.  This is different from the 
case of GI interaction \cite{pascalutsa_1999,Clymton:2017nvp}.

By using the propagator given in Eq.~(\ref{eq:propagator_pure}) and the interaction vertex factors
of Eqs.~(\ref{eq:piNDelta}) and (\ref{eq:gammaNDelta}) we may write the production amplitude as
\begin{eqnarray}
\mathcal{M}&=&\bar{u}_{N'} \,\Gamma_{\pi N \Delta}^{\rho\sigma}\, \frac{\Delta_{\rho\sigma\gamma\delta}}{s-m_\Delta^2 +i m_\Delta \Gamma_\Delta}\, \Gamma_{\gamma N \Delta}^{\gamma\delta}\,u_N \nonumber\\
&=& \bar{u}_{N'} \,\gamma_5 (p+k)_{\mu} q_\nu\, \mathcal{P}^{\mu\nu\alpha\beta}\, 
\Bigl[ G_1(k_\alpha\epsilon_\beta-\epsilon_\alpha k_\beta)\nonumber\\
&&+ G_2\,(\slashed{k}\epsilon_\alpha- k_\alpha\slashed{\epsilon})(p+k)_\beta\Bigr]\,u_N \, ,
\label{eq:production_amplitude}
\end{eqnarray}
where $m_\Delta$ and $\Gamma_\Delta$ are the mass and width of $\Delta$, respectively,
$s=(p+k)^2$, and we have used the relation 
\begin{eqnarray}
\mathcal{P}^{\rho \sigma \mu \nu}\Delta_{\rho\sigma\gamma\delta}
\mathcal{P}^{\alpha\beta\gamma\delta} = \mathcal{P}^{\mu\nu\alpha\beta} ~.
\end{eqnarray}
Furthermore, in Eq.~(\ref{eq:production_amplitude}) we have defined 
\begin{eqnarray}
G_1 &=& -\frac{g_1\,g_{\pi N\Delta}}{m_\Delta^2(s-m_\Delta^2 +i m_\Delta \Gamma_\Delta)} \, ,\\
G_2 &=& -\frac{g_2\,g_{\pi N\Delta}}{m_\Delta^3(s-m_\Delta^2 +i m_\Delta \Gamma_\Delta)} ~.
\end{eqnarray}
To calculate the total cross section we decompose the reaction amplitude ${\cal M}$
into the form functions $A_i$
\cite{Clymton:2017nvp}
\begin{eqnarray}
{\cal M} &=& {\bar u}_{N}(p') \sum_{i=1}^4 A_{i}(s,t,u)\, M_{i}\, u_N (p)~ ,
\label{eq:scattering-amplitudes-Mi}
\end{eqnarray}
with the gauge and Lorentz invariant matrices $M_i$ 
\begin{eqnarray}
\label{eq:M1}
M_{1} & = & \gamma_{5} \, \epsilon\!\!/ k\!\!\!/ ~ ,\\
\label{eq:M2}
M_{2} & = & 2\gamma_{5}\left(q \cdot \epsilon P \cdot k - q
\cdot k P \cdot \epsilon \right)~ ,\\
\label{eq:M3}
M_{3} & = & \gamma_{5} \left( q\cdot k \epsilon\!\!/ - q\cdot \epsilon 
k\!\!\!/ \right) ~ ,\\
\label{eq:M4}
M_{4} & = & i \epsilon_{\mu \nu \rho \sigma} \gamma^{\mu} q^{\nu}
\epsilon^{\rho} k^{\sigma}~ ,
\end{eqnarray}
where $P = \frac{1}{2}(p + p')$ and $\epsilon$ the photon polarization.
By performing the decomposition
we obtain
\begin{eqnarray}
A_1 &=& 2\left\{ k \cdot (q- p)-{\textstyle\frac{2}{3}}m_\pi^2+{\textstyle\frac{2}{3}}
  q\cdot(p+k)\right\}G_1 
\nonumber\\ &&+{\textstyle\frac{1}{3}}m_N\left(9p\cdot k+5m_\pi^2 -8 q\cdot(p+k)\right)G_2 \, ,\\
A_2 &=& 0 \, ,\\
A_3 &=& {\textstyle\frac{2}{3}} m_NG_1 \nonumber\\
    && + \Bigl\{s+{\textstyle\frac{2}{3}}m_\pi^2 
  - {\textstyle\frac{4}{3}}m_N^2 -3q\cdot(p+k)\Bigr\} G_2\, ,\\
A_4 &=& -{\textstyle\frac{4}{3}} m_NG_1 \nonumber\\
    && + \Bigl\{2s+{\textstyle\frac{2}{3}} m_\pi^2
    - {\textstyle\frac{4}{3}} m_N^2 -3q\cdot(p+k)\Bigr\}G_2 \, ,~~
\end{eqnarray}
from which we can calculate the total cross section~\cite{Knochlein:1995qz}.
Note that the form functions $A_i$ for non-ATS models can be found in the
previous works \cite{Clymton:2017nvp,david}.

\begin{figure}[t]
  \begin{center}
    \leavevmode
    \epsfig{figure=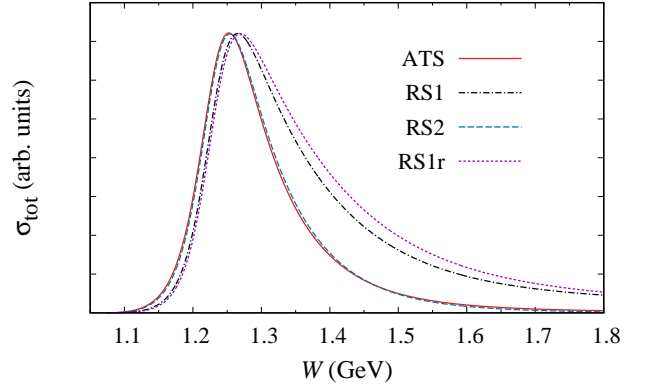,width=85mm}
    \caption{(Color online) Total cross sections of the $\pi N$ photoproduction
      calculated from the contribution of the $\Delta(1232)$ resonance 
      in arbitrary units according to the ATS prescription with consistent
      interaction Lagrangians (ATS), Rarita-Schwinger with GI interaction (RS1 
      \cite{pascalutsa_2001}), 
      Rarita-Schwinger with non-GI interaction (RS2 \cite{david}), and 
      Rarita-Schwinger with GI interaction with only two electromagnetic couplings
      (RS1r), as a function of the total c.m. energy $W$. Note that for the sake 
      of comparison the total cross section peaks are scaled to the same value.
      }
   \label{fig:contribution_gN} 
  \end{center}
\end{figure}

Different from pion scattering, in pion photoproduction there is only one 
hadronic vertex. Nevertheless, in the photoproduction the hadronic form 
factor still plays an important role to suppress the background contribution 
at high energies. Note that a very soft form factor leads to very strong 
suppression of the cross section. Although it produces an ideal resonance bump, the
resonance contribution could become very small and might distort the physics behind it.
On the other hand, a very hard form factor could fail to suppress the cross section
at high energies and, as a consequence, could fail to create the resonance bump.
Thus, as in the previous example we include the hadronic form factor given by 
Eq.~(\ref{eq:hff}), albeit with a different hadronic cutoff, i.e., $\Lambda=0.8$ GeV, 
to obtain reasonable values of photoproduction total cross section. 

In the present work we scale all peaks of the total cross sections to the same 
value. This is required merely for the sake of comparison, but we believe that this is 
still acceptable since we set all coupling constants to unity in the numerical calculation.
Moreover, in this visualization we merely want to see the structure of a resonance
produced by different representations of spin-3/2.
The result is shown in Fig.~\ref{fig:contribution_gN}, where we compare contribution of
the $\Delta(1232)$ resonance to the total cross section of pion photoproduction off
a nucleon according to the model of pure spin-3/2 with consistent interaction, the
RS model with the GI interaction \cite{pascalutsa_2001}, the RS model with non-GI 
interaction \cite{david}, and the modified RS with GI interaction \cite{Pascalutsa:2006up},
where in the latter we only use two electromagnetic couplings, instead of four 
as in the original version \cite{pascalutsa_1999}. Obviously all models 
exhibit a peak as the basic behavior required for a resonance. Surprisingly, the 
pure spin-3/2 with consistent interaction (ATS) and the Rarita-Schwinger with non-GI
interaction (RS2) prescriptions yield a similar structure. Presumably this is 
because the similar structure of the two models. The Rarita-Schwinger with 
GI interaction (RS1) indicates a larger background.
Previously, we suspect that this property originates from the larger number of coupling
constants used in this model. However, the use of only two of these coupling constants 
(RS1r), i.e., only $g_1$ and $g_2$, does not reduce the background. This is caused
by the destructive effect of the other two couplings, $g_3$ and $g_4$, that is missing
in the RS1r model. Nevertheless, the most important point to note here is that the
pure spin-3/2 model with consistent interaction produces the correct property of
resonance as in the conventional models. 

\section{Summary and Conclusion}
\label{sec:conclusion}

We have proposed the use of pure spin-3/2 propagator along with the 
consistent interaction Lagrangians in the phenomenological studies of 
nuclear and particle physics. To this end we employ the ATS representation 
to describe the corresponding projection operator. We have shown that the ATS 
formalism has a problem to exhibit the resonance behavior, unless the 
interaction Lagrangian is slightly modified, i.e., by replacing the gamma matrix 
with a partial derivative. However, the choice of a partial derivative seems 
to be arbitrary. To obtain a more systematic procedure and the correct degrees
of freedom we determine a number of constraints required by the interaction. 
In this work we give the simplest example of
consistent interactions that satisfy these constraints.
To visualize the result we apply the pure spin-3/2 propagator and consistent 
interactions to calculate the contribution of $\Delta(1232)$ resonance in the 
pion scattering and pion photoproduction off a nucleon. The obtained total cross 
sections in the two cases 
indicates that the pure spin-3/2 propagator with consistent 
interaction Lagrangians exhibits the required property of a resonance. Thus, 
we have proven that the proposed spin-3/2 propagator along with the consistent 
interaction Lagrangian can be directly used for phenomenological investigations 
in the realm of nuclear and particle physics.

\vspace{5mm}
\section{ACKNOWLEDGMENTS}

The authors acknowledge support from the Q1Q2 Grant of Universitas Indonesia, 
under contract No. NKB-0277/UN2.R3.1/HKP.05.00/2019.

\end{document}